**Traffic Wave Properties for Automated Vehicles During Traffic Oscillations via Analytical Approximations**

**Yang Zhou**
Assistant Professor
Zachry Department of Civil and Environmental Engineering
Texas A&M, College Station
199 Spence Street, College Station, TX, 77843
Email: yangzhou295@tamu.edu
Phone: (979) 458-6886

**Sixu Li**
Research Assistant
Department of Multi-disciplinary Engineering
Texas A&M, College Station
199 Spence Street, College Station, TX, 77843
Email: sixuli@tamu.edu

**Wissam Kontar**
Postdoctoral Research Associate
Department of Civil and Environmental Engineering
University of Wisconsin, Madison
1415 Engineering Drive
Madison, WI 53706
Email: kontar@wisc.edu

**Fan Pu**
Research Assistant
Zachry Department of Civil and Environmental Engineering
Texas A&M, College Station
199 Spence Street, College Station, TX, 77843
Email: fan-pu@tamu.edu

**Anupam Srivastava**
Researcher
Department of Civil and Environmental Engineering
University of Wisconsin, Madison
1415 Engineering Drive
Madison, WI 53706
Email: asrivastava2@wisc.edu

**Soyoung Ahn (Corresponding Author)**
Professor
Department of Civil and Environmental Engineering
University of Wisconsin, Madison
1415 Engineering Drive
Madison, WI 53706
Email: sue.ahn@wisc.edu
Phone: (608)265-9067

Word Count: 6606 words + 1 table (250 words per table) = 6,856 words

*Submitted [August 1,2024]*

**ABSTRACT**

This paper presents an analytical approximation framework to understand the dynamics of traffic wave propagation for Automated Vehicles (AVs) during traffic oscillations. The framework systematically unravels the intricate relationships between the longitudinal control model of the AVs and the properties of traffic waves. We apply Laplacian Transformation and Describing Function Analysis to mathematically derive the traffic wave properties of an AV in car-following scenarios. Further, we incorporate Newell's car-following model to determine the speed of the traffic waves. Our analysis extends to both homogenous and heterogenous traffic, systematically handling intra-heterogeneities and inter-heterogeneities in traffic wave propagation using the established analytical framework. We validate our approach via numerical simulations and show the connections between the AV control system and traffic wave properties. This research emphasizes the importance of rethinking our understanding of traffic wave properties when AVs are present in the traffic system.

**INTRODUCTION**

The fundamental understanding of traffic wave formation and propagation has long been the cornerstone of describing traffic dynamics. Traffic waves separate different traffic states in the time-space domain and signal transitions from one state to another. Given its paramount importance, extensive literature has explored the nature and properties of these waves for human driven vehicles (HDVs). Traffic waves have been famously modeled by kinematic wave models (*1–8*). This family of models assumes existence of a well-defined flow-density relationship (known as the "fundamental diagram") and flow conservation law, expressed as a partial differential equation. The solution to this problem provides traffic waves – orientation (forward, constant, or backward moving) and speed. In the original model by (*1, 4, 9*), traffic waves can vary depending on the initial and new traffic states, even within congestion. Newell's simplified kinematic wave model (*2, 3*) provides a simplified solution, where traffic waves in congestion are constant regardless of the traffic state. The kinematic wave models provide an invaluable tool to analyze congestion propagation and are often used in design of traffic control such as ramp metering and variable speed limit (*10–15*).

The properties of traffic waves in congestion are related to driver behavior, particularly car following (CF) behavior. However, for many CF models, a direct connection between CF dynamics and traffic wave properties is not trivial to establish, especially when the model is not parsimonious. A notable exception is Newell's simplified CF model (*16*) and its extensions (*17–22*). Newell postulated that the trajectory of a vehicle mirrors that of the leader but with a shift in time and space, corresponding to the driver's response time and minimum spacing, respectively. This shift thus represents how a disturbance travels as a wave from the leader to the follower. Since the response time and minimum spacing may vary by driver, a traffic wave propagates in a random walk fashion through vehicles, but without amplification or decay. The average wave speed quickly converges over vehicles (*23*). In Newell's model, the two parameters are assumed constant and independent of speed in congested traffic. Thus, the average wave speed (across vehicles) is also constant and corresponds to the congestion wave speed in the simplified kinematic wave theory. By no means, Newell's model is effective for capturing higher-order dynamic features, such as disturbance growth and traffic hysteresis. However, many empirical studies found some evidence that traffic waves propagate at approximately constant speed in congestion (*17*). While some variations in waves are often observed, attributable to the driver inter/intra heterogeneities (*18, 19, 24–26*), constant congestion wave speed is largely accepted and widely used in traffic science literature.

The rise of automated vehicles (AVs) brings forth a new era of traffic flow systems: mixed traffic comprising of both AVs and HDVs. This mixed traffic flow system challenges our conventional understanding of traffic flow due to the difference in behavior between AVs and HDVs. An essential aspect in understanding and modeling mixed traffic flow lies in understanding the properties of traffic waves specific to AVs. Compared to HDVs, AV CF control, also known as Adaptive Cruise Control (ACC), is envisioned to be more responsive and sensitive to traffic disturbances. Indeed, several experimental studies show that the CF behavior of AVs significantly differ from that of HDVs (*27, 28*). Further, higher-order features in oscillatory stop-and-go traffic can be more pronounced in AV traffic, depending on how the control is designed, which may violate the property of constant traffic wave speed (*29*). All these raise the question of if and how traffic wave properties differ for AVs under disturbances.

In a prior study by the authors (*30*), we analyze how the AV control structure and its parameter settings influence the response of AVs to traffic disturbances. While the work sheds light on the response of the AVs, a comprehensive understanding of the nature and properties of the traffic waves resulting from this response remains elusive thus far. Gaining understanding into these traffic waves is paramount for the development of effective control strategies for AVs and achieving desired traffic-level properties, such as string stability.

To this end, this paper aims to answer the following questions in the traffic oscillation scenario (with cyclic disturbances), where the high-order dynamics are more pronounced: (1) Is traffic wave within the congestion regime constant over time-space diagram, with the homogenous AV law? and (2) Under heterogenous AV CF law, what are the properties aligning with and differing from prevailing understandings? To comprehensively answer these questions, this paper establishes a theoretical foundation for comprehending traffic waves in AVs, uncovering the connections between AV control CF law and the fundamental properties of traffic waves. Due to the analytical tractability of linear AV law, we primarily focus on the linear AV car-following and further extend it to weak nonlinear scenarios such as the speed boundaries. As far as the authors are aware, this is the first paper to systematically discuss the AV waves in a theoretical and analytical way.

The remainder of this paper is organized as follows: **Section 2** presents a theoretical derivation of traffic waves for a homogeneous platoon with a single frequency oscillation, serving as the foundation for our framework. Building upon this foundation, **Section 3** extends the framework to heterogeneous traffic flow and generalizes the derivation by considering compounding frequency oscillations. To validate the framework and showcase physical understanding of traffic wave properties in AVs, **Section 4** presents numerical experiments.

## TRAFFIC WAVE THEORETICAL APPROXIMATION FOR HOMOGENOUS PLATOON WITH SINGLE FREQUENCY OSCILLATION

This section first provides a linear/linearized AV CF scenario example to illustrate our derivation framework for homogenous platoon. The example provides the wave speed approximation given the steady state oscillatory traffic condition by assuming that the speed oscillates with a predominant frequency. Based on that, we extend the framework for linear AV CF with weak nonlinearities, to make the framework more realistic.

Without loss of generality, the leading vehicle, indexed by vehicle 0, travels with a trajectory $p_0$ with two components: a nominal trajectory $\overline{p_0}$ and an oscillatory trajectory $\widehat{p_0}$. $\overline{p_0}$ travels with an equilibrium speed $v_e$, whereas $\widehat{p_0}$ follows a sinusoidal wave $\widehat{p_0}(t) = A\sin(\omega_m t)$, where $A_m, \omega_m$ denotes the oscillation magnitude and frequency, respectively. Following the kinematic motion law, we can readily have that, $v_0 = v_e + \widehat{v_0}$, and $\widehat{v_0}$, the oscillatory component for speed, follows $\widehat{v_0} = A\omega_m \cos(\omega_m t)$, and the acceleration follows $a_0 = -A\omega_m^2 \sin(\omega_m t)$.

Given a general automated vehicle linear/linearized CF controller, given as in Eq.(1), which states that the desired acceleration of vehicle 1, $u_1(t)$, follows the nonlinear mapping function $f$ with respect to the deviation from equilibrium spacing, leading vehicle speed $v_0$, and following vehicle speed $v_1$:

$$u_1(t) = f(\Delta p_1(t) - \Delta p_e, v_0(t), v_1(t)) \tag{1}$$

where $\Delta p_1(t) = p_0(t) - p_1(t)$ denoting the spacing, $\Delta p_e$ denotes the equilibrium spacing, which is a constant given a fixed $v_e$. Considering the linear controller or linearized control and an actuation time-lag of the vehicle dynamics, the actual acceleration of vehicle 1, $a_1(t)$ can be represented as:

$$\dot{a}_1(t) = -\frac{1}{\phi}a_1(t) + \frac{1}{\phi}(f_{p,1} \times (\Delta p_1(t) - \Delta p_e) + f_{v,1} \times v_0(t) + f_{v,0} \times v_1(t)) \qquad (2)$$

where $\phi$ represents the actuation time-lag, $f_{p,1}, f_{v,1}, f_{v,0}$ are corresponding linear/linearized feedback coefficient with respect to $\Delta p_1(t) - \Delta p_e$, $v_0(t)$ and $v_1(t)$ given as:

$$f_{p,1} = \frac{\partial f}{\partial(\Delta p_1 - \Delta p_e)}|_{\Delta p_1 = \Delta p_e} \qquad (3\text{-a})$$

$$f_{v,1} = \frac{\partial f}{\partial v_1}|_{v_1 = v_e} \qquad (3\text{-b})$$

$$f_{v,0} = \frac{\partial f}{\partial v_0}|_{v_0 = v_e} \qquad (3\text{-c})$$

By conducting the Laplacian transformation on Eq. (2), we can have:

$$G(s) = \frac{\widehat{p_0}(s)}{\widehat{p_1}(s)} = \frac{\widehat{v_0}(s)}{\widehat{v_1}(s)} = \frac{-f_{p,1} + f_{v,0}s}{\phi s^3 + s^2 - f_{v,1}s - f_{p,1}} \qquad (4)$$

As widely applied in control, we let $s = j\omega$, $j$ is the imaginary unity and $\omega \in (0, +\infty)$ is the frequency in rad, which means that frequency domain analysis mainly focused on steady oscillatory scenario. Specifically, $\omega = 2\pi f$. $f$ is the frequency in Hz. The transfer function $G(j\omega)$ can be further represented as $G(j\omega) = |G(j\omega)|\angle G(j\omega) = |G(j\omega)|e^{\angle G(j\omega)}$, $|G(j\omega)|$ is the norm function with respect to each frequency $\omega$, representing the disturbance (i.e. $\widehat{p_0}(s)$ or $\widehat{v_0}(s)$) amplification ratio, and $\angle G(j\omega)$ is the angle function with respect to each frequency $\omega$, representing the phase shift. Note that, $\angle G(j\omega)$ is directly related to the wave speed calculation. In general,

$$G(j\omega) = \frac{-f_{p,1} + j\omega f_{v,0}}{-j\omega^3\phi - \omega^2 - j\omega f_{v,1} - f_{p,1}} = \frac{AC + BD}{C^2 + D^2} - i\frac{CB - AD}{C^2 + D^2} \qquad (5\text{-a})$$

where,

$$A = -f_{p,1} \qquad (5\text{-b})$$
$$B = \omega f_{v,0} \qquad (5\text{-c})$$
$$C = -\omega^2 - f_{p,1} \qquad (5\text{-d})$$
$$D = -\omega^3\phi - \omega f_{v,1} \qquad (5\text{-e})$$

Then, we can further have:

$$|G(j\omega)| = \sqrt{\frac{A^2 + B^2}{C^2 + D^2}} \qquad (6\text{-a})$$

$$\angle G(j\omega) = \arctan\left(\frac{CB - AD}{AC + BD}\right) \qquad (6\text{-b})$$

Given simplified Newell's model as an example, $G(j\omega) = e^{-j\varphi}$, and $|G(j\omega)| = 1$, $\angle G(j\omega) = -\varphi$, where $\varphi$ is the time displacement of the Newell's model times the frequency $\omega$.

Based on Eq. (4), we have:

$$\widehat{p_1}(t) = A_m|G(j\omega_m)|\sin(\omega_m t + \angle G(j\omega_m)) \qquad (7)$$

For the CF control, the controller is usually designed with local (individual stability), which ensures that $\overline{p_1}(t) = \overline{p_0}(0) + s_e + v_e t$. By that we can analytically derive $p_1(t)$ as:

$$p_1(t) = \overline{p_0}(0) - s_e + v_e t + A_m|G(j\omega)|\sin(\omega_m t + \angle G(j\omega)) \qquad (8)$$

With the analytical forms given by Eq.(8), we can mathematically derive the wave speed function between vehicle 0 and vehicle 1, based on the generalized Newell's law shown by Fig.1.

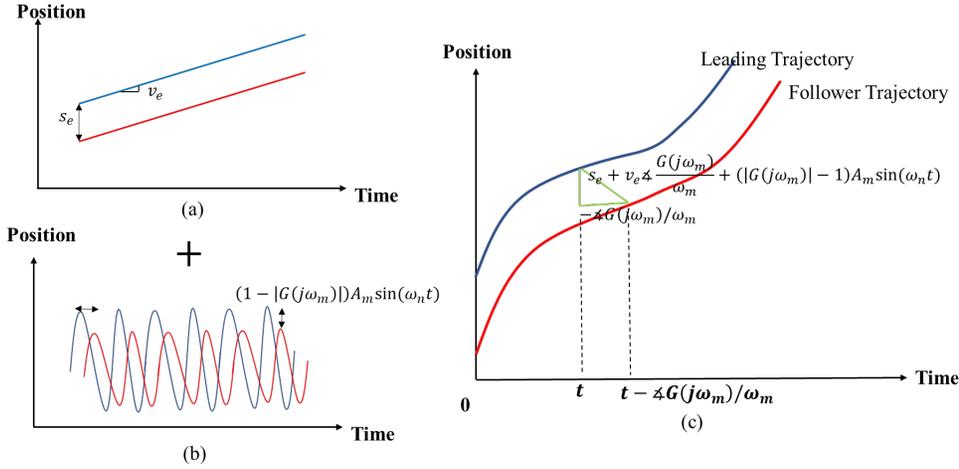

**Fig. 1 Wave Illustration (a) Nominal Trajectory; (b) Oscillatory Trajectory; (c) Original Trajectory**

As can be found that from Fig. 1(a), the nominal follower trajectory is parallel to nominal leading trajectory, with a vertical shift of equilibrium spacing $s_e$. Whereas, the oscillatory trajectory plays a vital role in terms of response time, with a shift $-\angle G(j\omega)$ shown by Fig. (b). By the wave propagation routine practice as Fig. (3), we can find shifted distance $h_{0-1}(t)$ follows:

$$h_{0-1}(t) = s_e + v_e \angle G(j\omega_m)/\omega_m + (1 - |G(j\omega_m)|)A_m \sin(\omega_m t) \tag{9}$$

Therefore, the wave speed follows:

$$W_{0-1}(t) = \frac{h_{0-1}(t)}{-\angle G(j\omega_m)/\omega_m} = \frac{\omega_m(s_e + (1-|G(j\omega_m)|)A_m \sin(\omega_m t))}{-\angle G(j\omega_m)} - v_e \tag{10}$$

As can be found from Eq. (10). The wave speed greatly depends on the equilibrium state (i.e., $\Delta p_e$, $v_e$) and oscillation features (i.e., $\omega_m$, $A_m$). By taking the closer look of $|G(j\omega_m)|$ and $\angle G(j\omega_m)$, we can also find that the control gains (i.e., $f_{p,i}, f_{v,i}, f_{v,i-1}$), and the measurement method or measurement timing (i.e., $t$) also play a vital role, which exhibits higher-order dynamic features for traffic wave than traditional approach. Further, we can find out that, the wave speed can be both positive and negative depending on the sign of right hand side of Eq. (10). To build the connection with traditional approach, we present **Remark 1** emphasizing the implications of traffic wave derived above.

***Remark 1:*** Differing from traditional analysis, where the wave analysis is only conducted when the speed changes most (Chen et al., 2012), our formulation is continuous making the analysis more temporally-spatially descriptive. If we allow $\omega t = \pm\frac{\pi}{2}$, i.e., $t = \pm\frac{\pi}{2\omega}$, we can also analyze the wave as traditional fashion (i.e., when acceleration/deceleration are at maximum deviation).

Based on Eq. (10), we can further analyze the properties of traffic wave speed for homogenous traffic as ***Proposition 1*** and ***2***.

***Proposition 1 (Wave speed intra-heterogeneity):*** *The wave speed for the CF pair is intra-heterogeneous under oscillation of a single wave component.*

*Proof:* Given $\forall t_1, t_2 > 0$, and $t_1 \neq t_2$, we compute the wave speed difference

$$W_{0-1}(t_1) - W_{0-1}(t_2) = \frac{\omega_m(|G(j\omega_m)|-1)A_m}{\measuredangle G(j\omega_m)}[\sin(\omega_n t_1) - \sin(\omega_n t_2)] \tag{11}$$

$W_{0-1}(t_1) - W_{0-1}(t_2) = 0$ only holds when $t_1 = \frac{2k\pi}{\omega} + t_2$ or $t_1 = \frac{2k\pi}{\omega} + \frac{\pi}{2} - t_2$ for $k = 0, \pm 1, \pm 2 \ldots$

*Though there are some special conditions that the wave speed are equal due to the cyclic movement, the wave speed is intra-heterogeneous in general.*

$$Q.E.D$$

We further generalize the derivation for a homogenous platoon consisting of $N$ vehicles. By Eq. (4), it is natural to have:

$$G(s) = \frac{\hat{p}_{i-1}(s)}{\hat{p}_i(s)} = \frac{\hat{v}_{i-1}(s)}{\hat{v}_i(s)} = \frac{a_{i-1}(s)}{a_i(s)} \qquad \forall i = 0,1,2 \ldots, N-1 \tag{12}$$

Based on that, we can mathematically derive $\hat{p}_i(s)$ similar to Eq. (7) have:

$$\hat{p}_i(t) = |G(j\omega_m)|^i e^{j\measuredangle G(j\omega)} A_m \sin(\omega_m t) = A_m |G(j\omega_m)|^i \sin(\omega_m t + i\measuredangle G(j\omega)) \tag{13}$$

Similarly, we can derive $p_i(t)$ and $W_{(i-1)-i}(t)$ as below:

$$p_i(t) = \bar{p}_i(t) + \hat{p}_i(t) = \overline{p_0}(0) - is_e + v_e t + A|G(j\omega)|^i \sin(\omega t + i\measuredangle G(j\omega)) \tag{14}$$

Considering the wave propagation path, as Fig. 2, we derive the wave generated by vehicle 0 at time $t$ that will travel to $t + i\measuredangle G(j\omega_m)$, and correspondingly we can find the shifted distance $h_{(i-1)-i}(t)$ as:

$$h_{(i-1)-i}(t) = s_e + v_e \measuredangle G(j\omega_m) + |G(j\omega_m)|^{i-1}(1 - |G(j\omega_m)|)A_m \sin(\omega_n t) \tag{15}$$

Further, the wave speed between vehicle $i-1$ and vehicle $i$ follows:

$$W_{(i-1)-i}(t) = \frac{h_{(i-1)-i}(t)}{-\measuredangle G(j\omega_m)/\omega_m} = \frac{\omega_m(s_e + |G(j\omega_m)|^{i-1}(|1 - G(j\omega_m)|)A_m \sin(\omega_m t))}{-\measuredangle G(j\omega_m)} + v_e \tag{16}$$

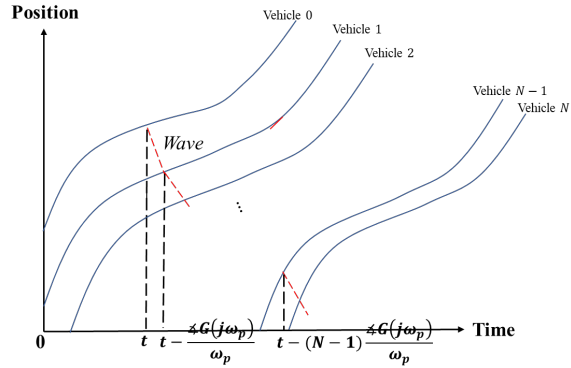

**Fig. 2 Wave Propagation Through Vehicle Strings**

Based on Eq. (16), we can conclude the wave speed inter-heterogeneity property (i.e., wave speed is different from vehicle to vehicle).

***Proposition 2 (Wave speed inter-heterogeneity):*** *The wave speed for the CF pair is in general inter-heterogeneous under oscillation of a single wave component.*

*Proof:*
To prove the inter-heterogeneity, we need to prove that $W_{(i-1)-i}(t) \neq W_{i-(i+1)}(t)$

$$W_{(i-1)-i}(t) - W_{i-(i+1)}(t) = -\frac{\omega_m |G(j\omega_m)|^{i-1}(|G(j\omega_m)|-1)^2}{\angle G(j\omega_m)} A_m \sin(\omega_m t) \quad (17)$$

The only condition to ensure the $W_{(i-1)-i}(t) - W_{i-(i+1)} = 0$ is $|G(j\omega)| = 1$, and $\omega_m t = 2k\pi \pm \pi$. Whereas in general, $W_{(i-1)-i}(t) - W_{i-(i+1)} \neq 0$.

$$Q.E.D$$

Further as can be found by Eq. (16), when the platoon length $N + 1$ is big, we can have two cases of interests:

### Case 1: Disturbance Dampening Case (i.e. $|G(j\omega_m)| < 1$)

When $|G(j\omega_m)| < 1$ and platoon length is long enough, we have the $p_N(t) = \bar{p}_i(t) + \hat{p}_i(s) = \overline{p_0}(0) - Ns_e + v_e t + A|G(j\omega_m)|^N \sin(\omega_m t + i\angle G(j\omega_m))$, and $|G(j\omega_m)|^N \to 0$. $p_N(t) \approx \overline{p_0}(0) - Ns_e + v_e t$, which means that the last vehicle almost travels at the constant speed $v_e$ and is not impacted by the disturbance. In this case, there is no wave speed after vehicle N, since vehicles after vehicle N all travels at constant speed.

### Case 2: Disturbance Amplification case (i.e., $|G(j\omega_m)| > 1$)

When $|G(j\omega_m)| > 1$ and platoon length is long enough, we have $p_N(t)$ oscillates between $-\infty$ to $\infty$. This is unreasonable since the linear law does not consider boundary conditions of traffic flow (such as the speed cannot violate the speed limit (denoted as $v_e$) and be negative). The boundary condition makes the linear control law not linear anymore, in which $G(j\omega_m)$ does not hold to describe the disturbance propagation.

To solve the boundary issue systematically, we apply the describing function analysis based method (Li and Ouyang, 2013) to theoretically derive the nonlinear transfer function $|G_{nl}(s)|$ between $\hat{p}_{i-1}(s)$ and $\hat{p}_i(s)$, where the subscript $nl$ denotes nonlinear. In our case, we consider the case that vehicles' speed is within the range $[0, v_{free}]$, which means $\widehat{v_0}(s)$ should be within the range $[-v_e, v_{free} - v_e]$.

Given an input signal $I(t) = A_m \sin(\omega_m t)$. The core idea is to decompose the output signal $O(t)$ by Fourier Transformation as:

$$O(t) = Y_{1,1}\sin(\omega_m t) + Y_{1,2}\cos(\omega_m t) + Y_{2,1}\sin(2\omega_m t) + Y_{2,1}\cos(2\omega_m t) + \cdots \quad (17)$$

By assuming the low-pass property of $|G_{nl}(s)|$, $O(t)$ can be approximated as:

$$O(t) \approx Y_{1,1}\sin(\omega_m t) + Y_{1,2}\cos(\omega_m t) \quad (18)$$

where

$$Y_{1,1} = \frac{1}{\pi}\int_{-\theta}^{2\pi-\theta} O(t)\sin(\omega_m t)\, d(\omega_m t) \quad (19)$$

$$Y_{1,2} = \frac{1}{\pi}\int_{-\theta}^{2\pi-\theta} O(t)\cos(\omega_m t)\, d(\omega_m t) \quad (20)$$

Based on Eq. (19) and Eq. (20), we can have the describing function, $H_{nl}(j\omega) \approx \frac{I(j\omega)}{O(j\omega)}$, for which:

$$G_{nl}(j\omega_m) = |G_{nl}(j\omega_m)| \sphericalangle G_{nl}(j\omega_m) \qquad (21)$$

where,

$$|G_{nl}(j\omega_m)| = \frac{\sqrt{Y_{1,1}^2 + Y_{1,2}^2}}{A_m} \qquad (22)$$

$$\sphericalangle G_{nl}(j\omega_m) = arctan(\frac{Y_{1,2}}{Y_{1,1}}) \qquad (23)$$

Specifically, in our case we are interested in $G_{nl}(j\omega_m) \approx \frac{\hat{p}_{i-1}(s)}{\hat{p}_i(s)} = \frac{\hat{v}_{i-1}(s)}{\hat{v}_i(s)}$, and we let $\hat{v}_{i-1}(t)$ as $I(t)$ and $\hat{v}_i(t)$ as $O(t)$. Following the procedure as Eqs. (17-23), we can find there are four cases of active boundaries depending on $B$, which is illustrated by Fig. 3.

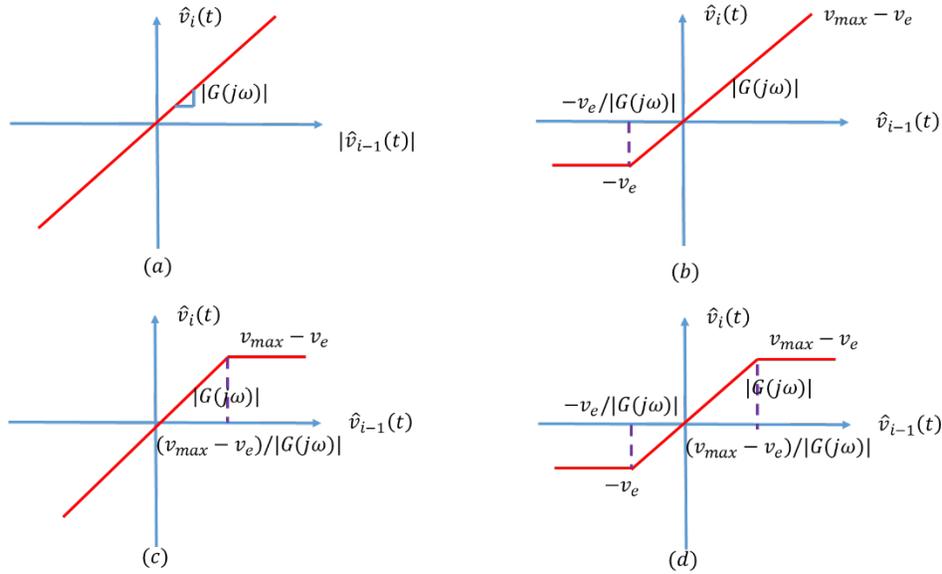

**Fig. 3** $\hat{v}_{i-1}(t)$ $vs.$ $\hat{v}_i(t)$ under different scenarios (a) Case 1; (b) Case 2; (c) Case 3; (d) Case 4

**Case 1** (Inactive Boundary): $|G_{nl}(j\omega_m)|A_m \geq v_e$, and $|G_{nl}(j\omega_m)|A_m \geq v_{free} - v_e$.

In this case, the boundaries of speed are inactive, which makes $G_{nl}(j\omega_m) = G(j\omega_m)$.

**Case 2** (Boundary activeness by free flow speed limit): $v_{free} - v_e \leq |G_{nl}(j\omega_m)|B \leq v_e$

In this case, we let $\beta_1 = \sin^{-1}(-\frac{v_e}{G(j\omega_m)})$ have:

$$\hat{v}_i(t) = \begin{cases} G(j\omega)\hat{v}_{i-1}(\omega_m t + \sphericalangle G(j\omega)) \text{ if } -\beta_1 - \sphericalangle G(j\omega_m) < \omega_m t \leq \pi + \beta_1 - \sphericalangle G(j\omega_m) \\ -v_e \text{ if } \pi + \beta_1 - \sphericalangle G(j\omega_m) < \omega t \leq 2\pi - \beta_1 - \sphericalangle G(j\omega_m) \end{cases} \qquad (24)$$

And correspondingly, we have

$$|G_{nl}(j\omega_m)| = \frac{\sqrt{\left(1-\frac{v_e^2}{G^2}\right)(4v_e^2+A^2v_e^2)+A^2G^2\left(\frac{\pi}{2}-\sin^{-1}\left(\frac{v_e}{G}\right)\right)^2}}{A} \tag{25}$$

$$\sphericalangle G_{nl}(j\omega_m) = \arctan\left(\frac{Y_{1,2}}{Y_{1,1}}\right) \tag{26-a}$$

And

$$Y_{1,1} = 2v_e \cos(\theta)\left(1-\frac{v_e^2}{G^2}\right)^{\frac{1}{2}} + AG\cos(\theta)\left(\frac{\pi}{2}-\sin^{-1}\left(\frac{v_e}{G}\right)\right) + Av_e\cos(\theta)\left(1-\frac{v_e^2}{G^2}\right)^{\frac{1}{2}} \tag{26-b}$$

$$Y_{1,2} = 2v_e \sin(\theta)\left(1-\frac{v_e^2}{G^2}\right)^{\frac{1}{2}} + AG\sin(\theta)\left(\frac{\pi}{2}-\sin^{-1}\left(\frac{v_e}{G}\right)\right) + Av_e\sin(\theta)\left(1-\frac{v_e^2}{G^2}\right)^{\frac{1}{2}} \tag{26-c}$$

**Case 3** (Boundary activeness by positive speed): $v_e \leq |G_{nl}(j\omega)|A_m \leq v_{free} - v_e$. In this case, we let $\beta_2 = \frac{v_{free}-v_e}{G(j\omega)}$ have:

$$\hat{v}_i(t) = \begin{cases} |G(j\omega_m)|\hat{v}_{i-1}(\omega_m t + \sphericalangle G(j\omega_m)) & \text{if } -\sphericalangle G(j\omega_m) < \omega_m t \leq \beta_2 - \sphericalangle G(j\omega_m) \\ v_{free} - v_e & \text{if } \beta_2 - \sphericalangle G(j\omega_m) < \omega_m t \leq \pi - \beta_2 - \sphericalangle G(j\omega_m) \\ |G(j\omega_m)|\hat{v}_{i-1}(\omega t + \sphericalangle G(j\omega_m)) & \text{if } \pi - \beta_2 - \sphericalangle G(j\omega_m) < \omega_m t \leq 2\pi - \sphericalangle G(j\omega_m) \end{cases} \tag{27}$$

$$|G_{nl}(j\omega_m)| = \frac{\left(4(v_{free}-v_e)^2\left(1-\frac{(v_{free}-v_e)^2}{G(j\omega_m)^2}\right)+A^2\left[2G(j\omega_m)\sin^{-1}\left(\frac{v_{free}-v_e}{G(j\omega_m)}\right)-2(v_{free}-v_e)\left(1-\frac{(v_{free}-v_e)^2}{G(j\omega_m)^2}\right)^{\frac{1}{2}}+G(j\omega_m)\pi\right]^2/4\right)^{1/2}}{A} \tag{28}$$

$$\sphericalangle G_{nl}(j\omega_m) = \arctan\left(\frac{Y_{1,2}}{Y_{1,1}}\right) \tag{29-a}$$

and

$$Y_{1,1} = 2(v_{free}-v_e)\cos(\theta)\left(1-\frac{(v_{free}-v_e)^2}{G^2}\right)^{\frac{1}{2}} + \{A\cos(\theta)[2G\sin^{-1}\left(\frac{v_{free}-v_e}{G}\right)-2(v_{free}-v_e)\left(1-\frac{(v_{free}-v_e)^2}{G^2}\right)^{\frac{1}{2}} + G\pi\}/2 \tag{29-b}$$

$$Y_{1,2} = 2(v_{free}-v_e)\sin(\theta)\left(1-\frac{(v_{free}-v_e)^2}{G^2}\right)^{\frac{1}{2}} + \{A\sin(\theta)[2G\sin^{-1}\left(\frac{v_{free}-v_e}{G}\right)-2(v_{free}-v_e)\left(1-\frac{(v_{free}-v_e)^2}{G^2}\right)^{\frac{1}{2}} + G\pi]\}/2 \tag{29-c}$$

**Case 4** (Boundaries activeness by positive speed and free flow speed limit): $\max(v_e, v_{free} - v_e) \leq |G_{nl}(j\omega)|A_m$

$$\hat{v}_i(t) = \begin{cases} |G(j\omega_m)|\hat{v}_{i-1}(\omega t + \sphericalangle G(j\omega_m)) & \text{if } -\beta_1 - \sphericalangle G(j\omega_m) < \omega_m t \leq \beta_2 - \sphericalangle G(j\omega_m) \\ v_{free} - v_e & \text{if } \beta_2 - \sphericalangle G(j\omega_m) < \omega_m t \leq \pi - \beta_2 - \sphericalangle G(j\omega_m) \\ |G(j\omega_m)|\hat{v}_{i-1}(\omega t + \sphericalangle G(j\omega_m)) & \text{if } \pi - \beta_2 - \sphericalangle G(j\omega_m) < \omega t \leq \pi + \beta_1 - \sphericalangle G(j\omega_m) \\ -v_e & \text{if } \pi + \beta_1 - \sphericalangle G(j\omega_m) < \omega_m t \leq 2\pi - \beta_1 - \sphericalangle G(j\omega_m) \end{cases} \quad (30)$$

$$|G_{nl}(j\omega)| = \frac{\sqrt{Y_{1,1}^2 + Y_{1,2}^2}}{A} \tag{31}$$

$$\sphericalangle G_{nl}(j\omega) = \arctan\left(\frac{Y_{1,2}}{Y_{1,1}}\right) \tag{32-a}$$

And
$$Y_{1,1} = 2(v_{free} - v_e)\cos(\theta) + 2v_e \cos(\beta_1)\cos(\theta) + \{AG(j\omega)\cos(\theta)[2\beta_1 + 2\beta_2 - \sin(2\beta_1) - \sin(2\beta_2)]\}/2 \quad (32\text{-b})$$

$$Y_{1,2} = 2(v_{free} - v_e)\sin(\theta) + 2v_e \cos(\beta_1)\sin(\theta) + \{AG(j\omega)\sin(\theta)[2\beta_1 + 2\beta_2 - \sin(2\beta_1) - \sin(2\beta_2)]\}/2 \quad (32\text{-c})$$

Note that, there are other potential boundaries such as acceleration/deceleration limit, that can be analyzed in a similar fashion. The corresponding transfer function rendered by other boundaries can also be derived based on the above framework built on DFA. Since this is not the main focus of this paper, we will leave the derivations of different boundaries' cases for readers with interest.

**EXTENSION DERIVATIONS AND PROPERTY ANALYSIS**
In this section, we further extend the above analysis by considering inter-vehicle controller heterogeneities. Based on that, we will further discuss and approximate the case of compounding traffic oscillations.

**Heterogeneous Controller Case**
The inter-vehicle controller heterogeneity is defined on the basis of Eqs.(1) and (3). If there exist a parameter among $f_p, f_v, f_{v,l}, \phi, \Delta p_e$, as well as nonlinearity boundaries that differs from one vehicle to another, we denote that inter-vehicle controller heterogeneity exists. The heterogeneities of parameters make $G_{nl}(j\omega_m)$ vary from each vehicle $i$, and we denote it as $G_i^{nl}(j\omega)$. Similar to Eq. (14), we can naturally find the relationship between the oscillation component:

$$G_i^{nl}(j\omega_m) = \frac{\hat{p}_{i-1}(z)}{\hat{p}_i(z)} = \frac{\hat{v}_{i-1}(z)}{\hat{v}_i(z)} = \frac{a_{i-1}(z)}{a_i(z)} \qquad \forall i = 0,1,2\ldots, N-1 \tag{33}$$

Which gives that:
$$\hat{p}_i(t) = A_m \prod_{h=1}^{i} |G_i^{nl}(j\omega_m)| \sin\left(\omega_m t + \sum_{h=1}^{i} \sphericalangle G_i^{nl}(j\omega_m)\right) \tag{34}$$

Accordingly, we have the position of vehicle $i$ follows:
$$p_i(t) = \overline{p_0}(0) - is_e + v_e t + A_m \prod_{h=1}^{i} |G_h^{nl}(j\omega_m)| \sin\left(\omega_m t + \sum_{h=1}^{i} \sphericalangle G_i^{nl}(j\omega_m)\right) \tag{35}$$

Similar to Eq. (16), the wave speed is
$$W_{(i-1)-i}(t) = \frac{\omega_m h_{(i-1)-i}(t)}{-\sphericalangle G(j\omega_m)} = \frac{\left(s_{e,i} + \left[|G_i^{nl}(j\omega_m) - 1|A_m \prod_{h=1}^{i-1} |G_h^{nl}(j\omega_m)|\right]\sin(\omega_m t)\right)\omega_m}{-\sphericalangle G_i^{nl}(j\omega_m)} + v_e \tag{36}$$

As can be found by Eq.(36), the intra/inter wave speed as **Proposition 1 (intra)** and **Proposition 2 (inter)** still hold. We further prove the following properties given as **Proposition 3** and **Proposition 4**, based on Eq. (36).

***Proposition 3 (Traffic wave commutative property of a single frequency oscillation)**: For a single frequency oscillation with frequency $\omega_m$, if the nonlinearity boundaries are not touched, the average wave will not change for the heterogeneous platoon regardless of vehicle sequence.*

*Proof:*

*To show the traffic wave commutative property, we consider vehicle platoon of arbitrary length $N + 1$ and with different platoon sequence $\mathbb{S}_1$ and $\mathbb{S}_2$. The wave travel time for these two different sequences is presented as:*

$t_w(\mathbb{S}_1) = \omega_m \sum_{i \in \mathbb{S}_1} \sphericalangle G_i^{nl}(j\omega_m) = \omega_m \sum_{i=0}^{N} \sphericalangle G_i(j\omega_m)$

$t_w(\mathbb{S}_2) = \omega_m \sum_{i \in \mathbb{S}_2} \sphericalangle G_i^{nl}(j\omega_m) = \omega_m \sum_{i=0}^{N} \sphericalangle G_i(j\omega_m)$

*We can find that: $t_w(\mathbb{S}_1) = t_w(\mathbb{S}_2) = t_w(N)$*

*Similarly, we have:*

$h_{\mathbb{S}_1}(t) = \sum_{i \in \mathbb{S}_1} s_{e,i} + v_e \sum_{i \in \mathbb{S}_1} \sphericalangle G_i^{nl}(j\omega_m) + h_{0-N}(t)$
$\qquad = \sum_{i=1}^{N} s_{e,i} + v_e t_w(N) + A_m (\prod_{i=1}^{N} |G_i(j\omega_m)| - 1) \sin(\omega_n t)$

$h_{\mathbb{S}_2}(t) = \sum_{i \in \mathbb{S}_2} s_{e,i} + v_e \sum_{i \in \mathbb{S}_2} \sphericalangle G_i^{nl}(j\omega_m) + h_{0-N}(t)$
$\qquad = \sum_{i=1}^{N} s_{e,i} + v_e t_w(N) + A_m (\prod_{i=1}^{N} |G_i(j\omega_m)| - 1) \sin(\omega_n t) = h_{\mathbb{S}_1}(t)$

*Since $W_{\mathbb{S}_1}(t) = \frac{h_{\mathbb{S}_1}(t)}{t_w(N)}$ and $W_{\mathbb{S}_2}(t) = \frac{h_{\mathbb{S}_2}(t)}{t_w(N)}$, we can also find $W_{\mathbb{S}_1}(t) = W_{\mathbb{S}_2}(t)$*

$$Q.E.D$$

***Proposition 4 (Traffic wave associative property of a single frequency oscillation):** For a single frequency oscillation with frequency $\omega_m$, if the nonlinearity boundaries are not touched, the wave propagation equals to the vector sum of the wave between sub-platoon.*

*Proof:*

Suppose $\forall k \in [1,2,\dots N]$, which splits the whole platoon $\mathbb{N}$ into two sub-platoons $\mathbb{N}_1, \mathbb{N}_2$. $\mathbb{N}_1$ begins from vehicle 0 to $k$, and $\mathbb{N}_2$ begins from vehicle $k+1$ to $N$. We have that the wave travel time for $\mathbb{N}_1$, $t_w(\mathbb{S}_1)$, follows:

$t_w(\mathbb{N}_1) = \omega_m \sum_{i=1}^{k} \sphericalangle G_i^{nl}(j\omega_m) = \omega_m \sum_{i=1}^{k} \sphericalangle G_i(j\omega_m)$
$t_w(\mathbb{N}_2) = \omega_m \sum_{i=k+1}^{N} \sphericalangle G_i^{nl}(j\omega_m) = \omega_m \sum_{i=k+1}^{N} \sphericalangle G_i(j\omega_m)$

We can verify that:
$$t_w(\mathbb{N}_1) + t_w(\mathbb{N}_2) = t_w(N)$$

Similarly, we have:

$h_{\mathbb{N}_1}(t) = \sum_{i=1}^{k} s_{e,i} + v_e t_w(\mathbb{N}_1) + h_{0-k}(t)$
$\qquad = \sum_{i=1}^{k} s_{e,i} + v_e t_w(\mathbb{N}_1) + A_m (\prod_{i=1}^{k} |G_i(j\omega_m)| - 1) \sin(\omega_n t)$

$h_{\mathbb{N}_2}(t) = \sum_{i=k+1}^{N} s_{e,i} + v_e t_w(\mathbb{N}_2) + h_{(k+1)-N}(t)$
$\qquad = \sum_{i=k+1}^{N} s_{e,i} + v_e t_w(\mathbb{N}_2) + A_m (\prod_{i=k+1}^{N} |G_i(j\omega_m)| - \prod_{i=1}^{k} |G_i(j\omega_m)|) \sin(\omega_n t)$

$h_{\mathbb{N}_1}(t) + h_{\mathbb{N}_2}(t) = \sum_{i=k+1}^{N} s_{e,i} + v_e t_w(\mathbb{N}_2) + A_m (\prod_{i=1}^{N} |G_i(j\omega_m)| - 1) \sin(\omega_n t) = h_{\mathbb{N}}(t).$

Since the beginning points and ending points in both temporal domain and space domain are same with/out sub-platooning, the associative of wave propagation holds.
$$Q.E.D$$

**Compounding Oscillations Case**

To further make the analysis more general, wo consider the general oscillation case, where the position oscillation components follow $\widehat{p_0}(t) = \sum_{m=1}^{M} A_m \sin(\omega_m t + \phi_m)$. $M$ is the total number of oscillations components. Similar to single wave component case, we can further have $\widehat{v_0} = \sum_{m=1}^{M} A_m \omega_m \sin(\omega_m t)$, and $\overline{p_0}(t) = \overline{p_0}(0) + s_e + v_e t$.

By Eq.(14), we can also derive $\widehat{p_1}(t), \hat{p}_i(t), p_i(t)$ by additivity:
$$\widehat{p_1}(t) = \sum_{m=1}^{M} |G_1^{nl}(j\omega_m)| A_m \sin(\omega_m t + \sphericalangle G(j\omega_m) + \phi_m) \tag{37}$$
and for any vehicle $i$ in the platoon of length $N+1$
$$\hat{p}_i(t) = \sum_{m=1}^{M} \prod_{h=1}^{i} |G_h^{nl}(j\omega_m)| A_m \sin(\omega_m t + i\sphericalangle G_h^{nl}(j\omega_m) + \phi_m) \tag{38}$$
$$p_i(t) = \overline{p_0}(0) - is_e + v_e t$$
$$+ \sum_{m=1}^{M} \prod_{h=1}^{i} |G_h^{nl}(j\omega_m)| A_m \sin(\omega_m t + i\sphericalangle G_h^{nl}(j\omega_m) + \phi_m) \tag{39}$$

As can be found that when the vehicle controller is strictly string stable (i.e., $sup_m |G(j\omega_m)| < 1$), and $N$ is big enough, we can also reach the conclusion that $p_N(t)$ remains the equilibrium spacing. However, differed from single oscillation case, where we can make an exact theoretical derivation, the compounding oscillation case is much complex and we can only provide an approximation. As suggested by Thieman et al., (2005) and Zhou et al.,(2020), there is usually a predominant frequency, which means that $A_{i,p} \gg A_{i,m}$, for $1 \leq m \leq M$ and vehicle $i$. Under this assumption, we can also approximate the wave speed by only considering the predominant oscillation component. Note predominance frequency may also alter when the traffic wave propagates through vehicle string followed by **Proposition 5** exampled by a homogenous platoon.

***Proposition 5 (Predominant frequency inter-heterogeneity):*** *Given the leading vehicle position oscillation components follows $\widehat{p_0}(t) = \sum_{m=1}^{M} A_m \sin(\omega_m t)$, and there exists a $p$ such that $A_p \gg A_m$, $1 \leq \forall m \leq M$ for the vehicle 0. If there exist a $m$ such that $|G(j\omega_m)| > |G(j\omega_p)|$, we can find a vehicle number $n$ that the fundamental frequency changes given the homogenous platoon length is long enough.*
*Proof:*
*By Eq. (38), we have $\hat{p}_i(t) = \sum_{m=1}^{M} |G(j\omega_m)|^i A_m \sin(\omega_m t + i\sphericalangle G(j\omega_m) + \phi_m)$ whose oscillation intensity is $|G(j\omega_m)|^i A_m$ for frequency $\omega_m$. By letting $|G(j\omega_m)|^i A_m = |G(j\omega_k)|^i A_k$, we can find that $\left(\frac{|G(j\omega_m)|}{|G(j\omega_p)|}\right)^i = \frac{A_p}{A_m}$. By taking log on both sides, we have $i\log\left(\frac{|G(j\omega_m)|}{|G(j\omega_p)|}\right) = \log\left(\frac{A_p}{A_m}\right)$, which simplifies to $i = \log\left(\frac{A_k}{A_m}\right)/\log\left(\frac{|G(j\omega_m)|}{|G(j\omega_k)|}\right)$. For any vehicle $i > \left\lceil \log\left(\frac{A_k}{A_m}\right)/\log\left(\frac{|G(j\omega_m)|}{|G(j\omega_k)|}\right) \right\rceil$, we can find that oscillation intensity for frequency $\omega_m$ is larger than that for frequency $\omega_p$, which suggests the change of fundamental frequency.*
$$Q.E.D$$

With the potential predominant frequency inter-heterogeneity, we further an approximated $h_{(i-1)-i}(t)$.

$$h_{(i-1)-i}(t) \approx s_{e,i} + A_p|G_{i,p}(j\omega_p) - 1||\tilde{G}_{i,p}(j\omega_p)|\sin(\omega_p t + \phi_p) \tag{40}$$

Where:

$$|\tilde{G}_{i,p}(j\omega_p)| = \prod_{h=1}^{i}|G_h^{nl}(j\omega_p)| \tag{41}$$

$$|G_{i,p}(j\omega_p)| = |G_p(j\omega_p)| \tag{42}$$

and $p$ should satisfy:

$$p = argmax_m A_m \prod_{h=1}^{i}|G_h^{nl}(j\omega_p)| \tag{43}$$

Similarly:

$$W_{(i-1)-i}(t) \approx \frac{\omega_m h_{(i-1)-i}(t)}{\measuredangle G_{i,p}(j\omega_p)} \tag{44}$$

## NUMERICAL EXPERIMENT

In this section, we seek to explore the physical implications of the analytical formulations provided in this work. Specifically, we build upon the numerical experimentation shown in this section to reveal the intricate connections between the controller design (i.e., control parameters of the linear controller) and the traffic wave properties. Specifically, Section 4.1 gives the default experiment setup and an example to illustrate the intra/inter wave heterogeneity. Based on that, Section 4.2 further unveil each parameters impact on wave speed. Section 4.3, we further the string stability impact on the wave speed. Section 4.4 will delve into the nonlinearity impact.

### Experimental Setup

To investigate the impacts of control parameters on the traffic wave properties, we run an experimental simulation Specifically, we set the CF law of AVs as a widely adopted third-order linear feedback controller by Zhou et al., (2019) as an example due to its analytical form. Based on Eq. (3), we get $f_{p,i} = -k_s$, $f_{v,i} = -k_v - k_s\tau$, $f_{v,i-1} = k_v$, and $s_e(t) = v_l(t) \times \tau + s_0$, where $k_s$, $k_v$, $\tau$, and $s_0$ represent spacing feedback gain, speed deviation feedback gain, constant desired time gap, and standstill spacing, respectively. The corresponding transfer function is given as:

$$G(j\omega) = \frac{k_s + j\omega k_v}{-j\omega^3\phi - \omega^2 - j\omega(k_v + k_s\tau) + k_s} \tag{45}$$

Correspondingly, by Eqs. (6-a) and (6-b), we have:

$$|G(j\omega)| = \sqrt{\frac{k_s^2 + k_v^2\omega^2}{(\omega^2 - k_s)^2 + (\omega^3\phi + \omega(k_v + k_s\tau))^2}} \tag{46-a}$$

$$\measuredangle G(j\omega) = \arctan\left(\frac{\omega k_v(k_s - \omega^2) - k_s(\omega^3\phi + \omega(k_v + k_s\tau))}{k_s(k_s - \omega^2) + \omega k_v k_s(\omega^3\phi + \omega(k_v + k_s\tau))}\right) \tag{46-b}$$

One can clearly note that the control parameter setting $(k_s, k_v, \tau, \phi)$ yields an influence on the traffic wave propagated by the controller. In here this lies the complexities, whereby the overall control logic generates different response to traffic oscillations, and thus traffic wave properties. We note that the control gains $k_s$ and $k_v$ in essence govern the CF behavior of the AV by regulating the controller's response to a deviation from equilibrium spacing, and speed difference, respectively. In general, we note that the influence of these controller gains on the CF behavior of an AV can be found in our previous work (*30*).

However, their influence on the wave properties can be more complex.

Table 1 below shows the default values of control parameters in our experimental setup. To further provide a direct illustration on the wave propagation over the homogeneous linear, we set the speed oscillation amplitude of the leading as 15 $m/s$, with $\omega = 0.16\pi\ rad/s$.

**TABLE 1 Experimental Setup**

| Control Parameter | Default Value |
|---|---|
| $k_s$ | 1 |
| $k_v$ | 1 |
| $\tau$ | 1.2 (sec) |
| $\phi$ | 0.1 |

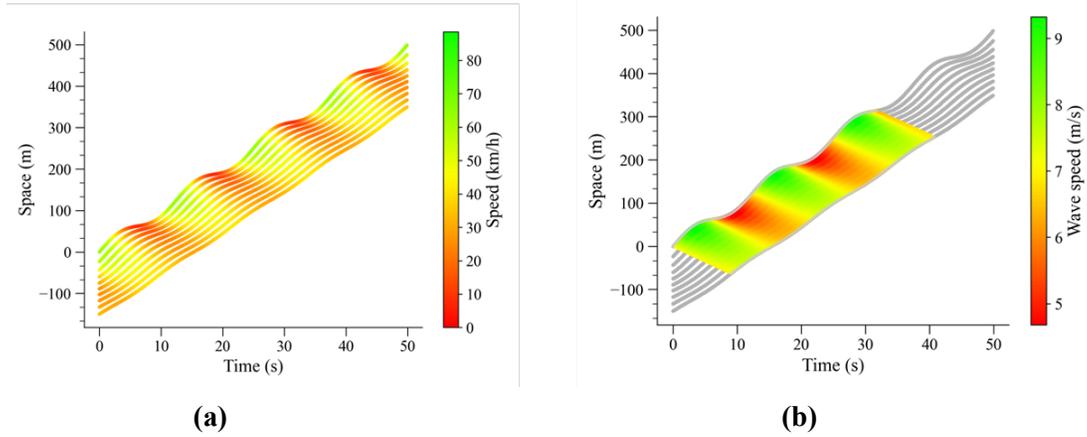

(a)            (b)

**Fig. 4** Time space diagram of homogenous AV (a) Vehicles Speed; (b) Wave Speed

As can be found from Fig.4 (a), given the single frequency oscillation, the traffic wave seems to be nonlinear in temporal and spatial domain. Given as closer look by Fig. 4(b), we can find that, the intra/inter heterogeneity do exist, which validate the propositions of this paper. To have a further understanding, the following experiments involves systematically varying individual control parameters while keeping others constant. The goal of this analysis is to explore the influence of each control parameter on wave, allowing us to gain insights into generic attributes that might help us design a controller for desired wave properties.

**Parametric Impact on Wave**

Given the default parameter of Table 1, we vary each parameter $k_s$, $k_v$, $\tau$ respectively to understanding the impact on the wave, and we use the wave speed between vehicle 0 and 1 as an example.

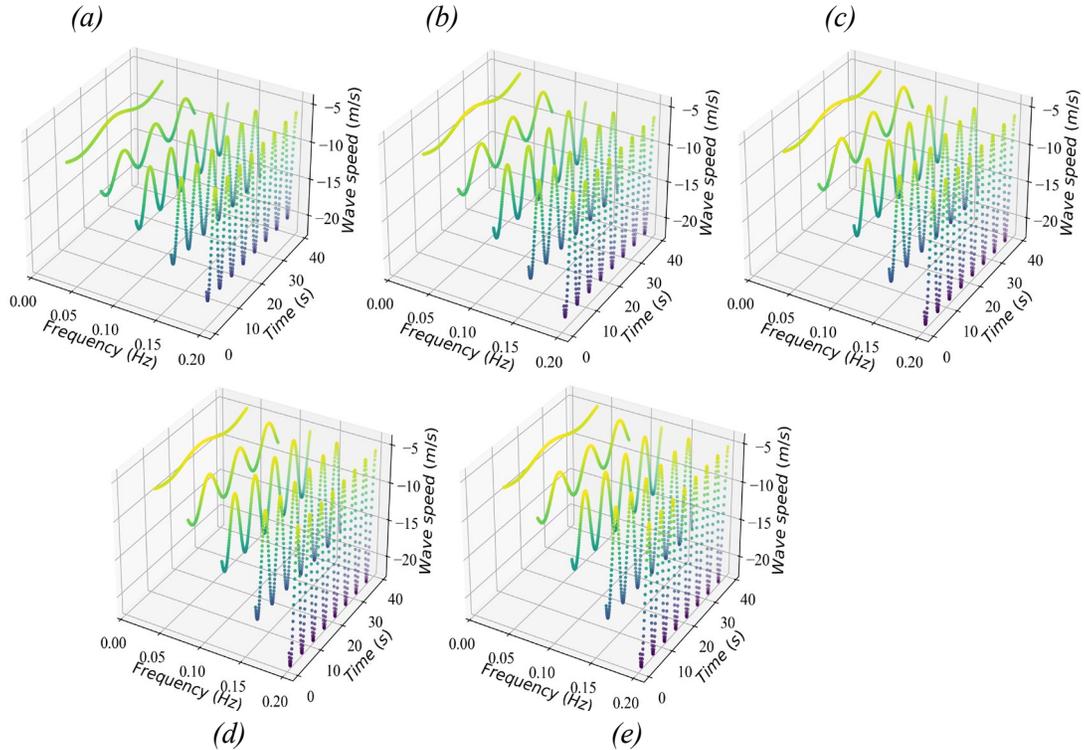

**Fig. 5** Wave speed over oscillations of different frequency (a) $k_s = 0.2$; (b) $k_s = 0.6$; (c) $k_S = 1$; $k_s = 1.4$

From Fig. 5 we can find that the evolutions of wave speed over time are periodic by changing the oscillation wave frequency. Increasing the spacing gain $k_s$, we observe an increase in the wave speeds (magnitude wise). This change is more evident on oscillation amplitude on wave speed and the amplitude increase over frequency. To provide further insights, the disturbance dampening ratio $|G(j\omega)|$ as well as the response time $-\angle G(j\omega)/\omega$ as given below:

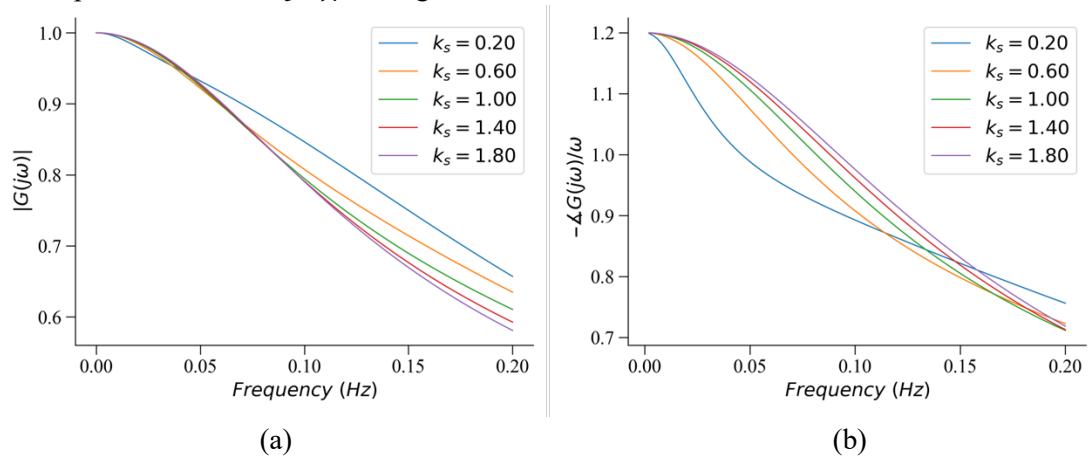

**Fig. 6** Disturbance amplification ratio and response time over different $k_s$ (a) Disturbance amplification ratio; (b) Response time

First, it is noted that the frequency of the traffic wave influences its dampening ratio. At higher frequency domains, making dampening of the oscillation more effective and reachable. When frequency is close to zero, $|G(j\omega)| \to 1$ and $-\sphericalangle G(j\omega)/\omega \to \tau$, which make the wave speed close to a constant, which is consistent with Simplified Newell's car following model. When frequency increase, at low frequencies $\leq 0.05hz$ there seems to be no significant influence of $k_s$ on dampening ratio. However, a slight benefit of higher $k_s$ on improving disturbance dampening is noted at frequencies $> 0.05hz$. On the other hand, the $k_s$ can provide the smaller response time over low frequency $\leq 0.1hz$, and the change flipped at high frequency $> 0.1hz$. By Eq. (10), we found the wave speed oscillation amplitude is linearly proportional to $|1-|G(j\omega)||$, and the inverse of response time. When frequency increase, the $|G(j\omega)|$ becomes and response time all become smaller, which induce to the increase of wave speed oscillation.

Under different $k_v$, we also have the similar finding the that when the frequency increase, the wave speed oscillation magnitude also increase. The oscillation on wave speed become more pronounced when $k_v$ increase.

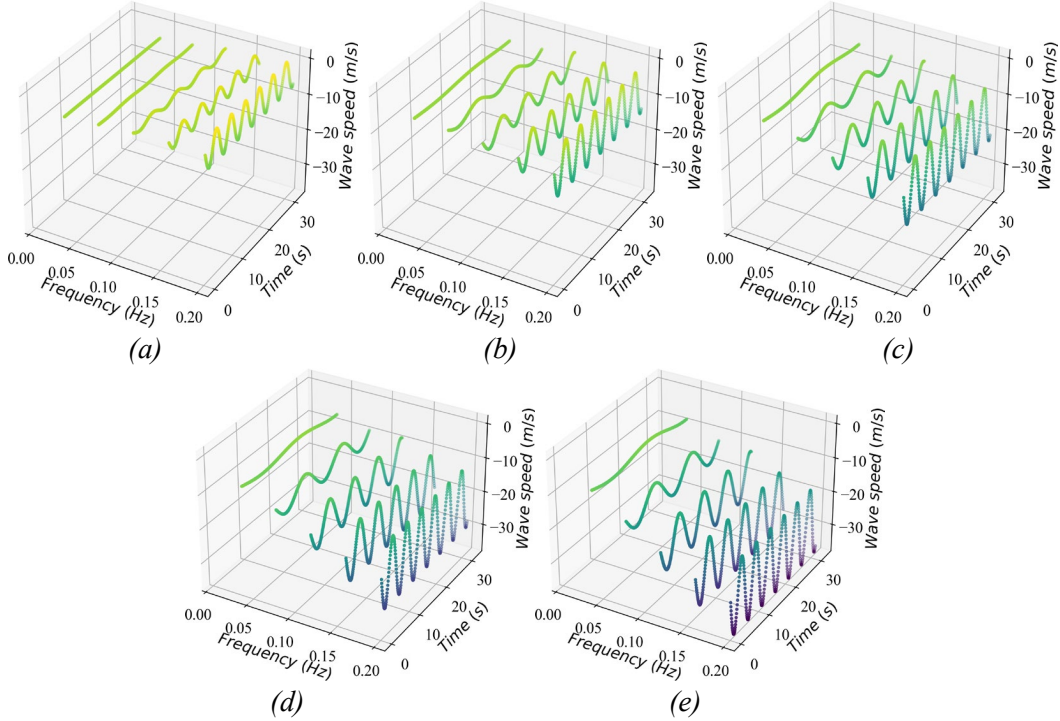

**Fig. 7** Wave speed over oscillations of different frequency (a) $k_v = 0.2$; (b) $k_v = 0.6$; (c) $k_v = 1$; $k_v = 1.4$

Similarly, the disturbance dampening ratio $|G(j\omega)|$ as well as the response time $-\sphericalangle G(j\omega)/\omega$ as given in Fig.8. From the results, we notice more complexity. In essence, the role of $k_v$ here is to regulate the magnitude of effort by which the controller matches the follower speed. This has direct implications on disturbance dampening and response time. For instance, when $k_v$ increase, a controller is sensitive and responsive to instantaneous speed change by the leader, the generated disturbance will be dampened in a more intensive and res. Though, this behavior seems to be more complex when examined through a frequency domain. At frequencies $\leq 0.1hz$, higher $k_v$ leads to significant benefits on dampening disturbances. Yet, at higher frequencies the role of $k_v$ is less effective and seems to flip a little. This might be attributed to the convoluted relation given in Eq. (46). Nevertheless, in general, the results still match the findings in Fig. 7.

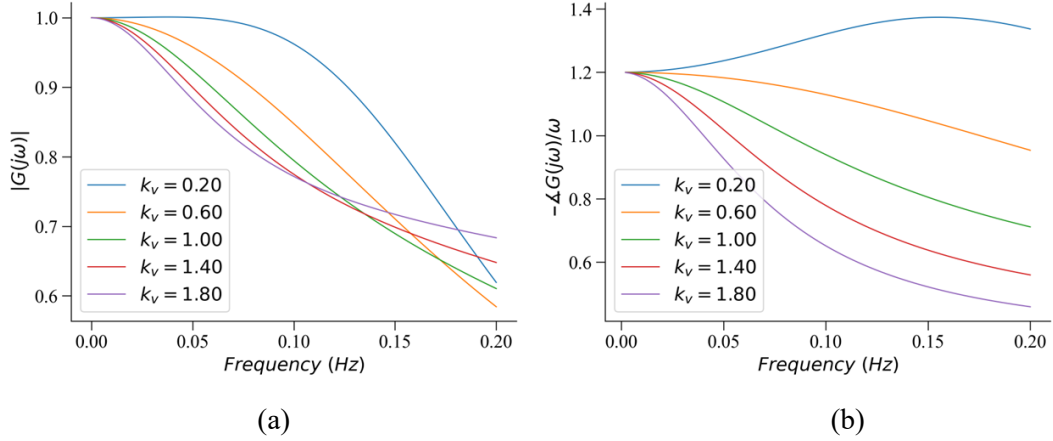

**Fig. 8** Disturbance amplification ratio and response time over different $k_v$ (a) Disturbance amplification ratio; (b) Response time

As for $\tau$, the wave is further given in Fig. 9 the frequency impact is similar to $k_s$ and $k_v$. Further, by larger $\tau$, the wave speed magnitude become smaller align with Simplified Newell's car following model. When $\tau$ increase, the amplification of wave speed oscillation also increase. To delve into that phenomenon, we further conduct the analysis on disturbance amplification ratio and response time as in Fig. 10.

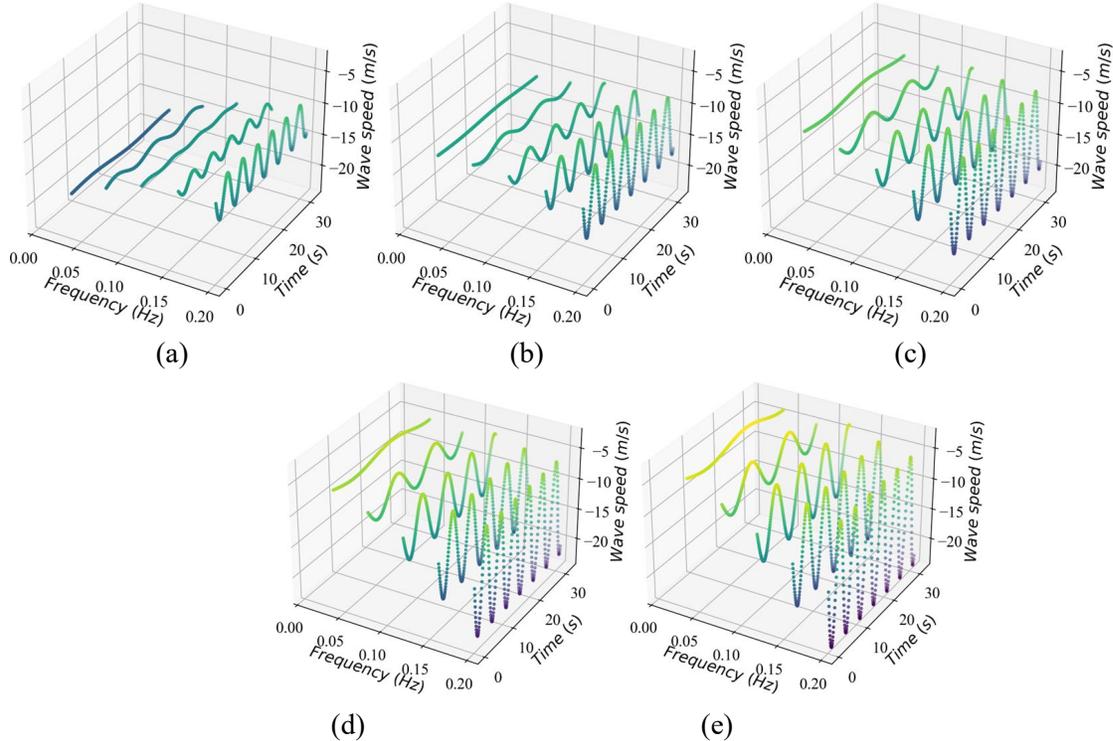

**Fig. 9** Wave speed over oscillations of different frequency (a) $\tau = 0.6$; (b) $\tau = 0.8$; (c) $\tau = 1$; (d) $\tau = 1.2$; (e) $\tau = 1.4$

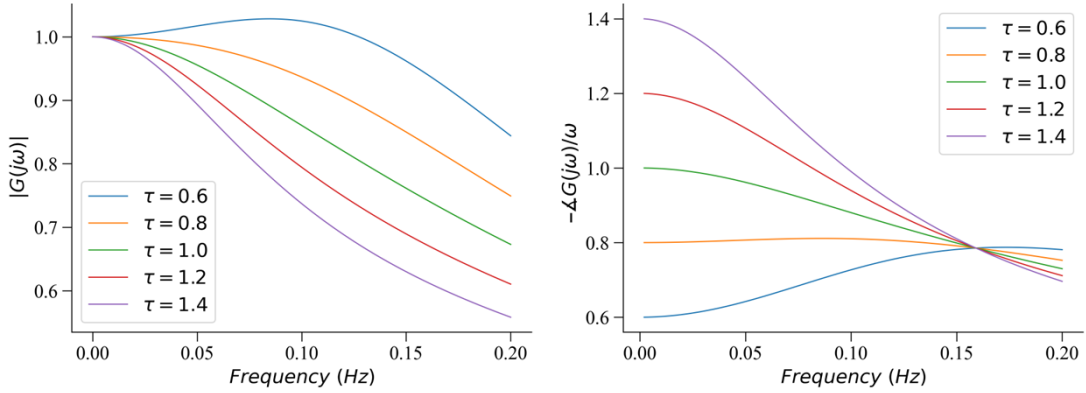

**Fig. 10** Disturbance amplification ratio and response time over different $\tau$ (a) Disturbance amplification ratio; (b) Response time

When $\tau$ increase, the dampening effect becomes more apparent, as it prompts a larger equilibrium spacing between the AV and its leader, effectively limiting the influence of disturbances. Correspondingly, the response time becomes larger, especially in the low frequency range $0.15 hz$, naturally by the definition of $\tau$. The trend flips at the high frequency range, in which scenario, the response is more dominated by high frequency features rather than the equilibrium time gap. By Eq. (10), the nonlinear relationship further induces to the wave speed amplification ratio increase when $\tau$ increase.

**String Stability Impact on Wave Propagation**

The aforementioned analysis stick to the analysis on leading vehicle and the first following vehicle. As can be found from Eq. (16), the wave propagation over vehicle platoon is impacted the $|G(j\omega)|^i$. By that we can envision that $|G(j\omega)| > 1$ (i.e., linear string instable) and $|G(j\omega)| \leq 1$ (i.e., linear string stable) can have significantly different trend over the platoon length $N$. To further analyze the wave propagation on the space, we focus on a homogenous AV platoon with five vehicles, where the speed oscillation amplitude and frequency of the leading vehicle are set as 15 $m/s$ and $f = 0.05\ Hz$, respectively. In the string instable case, we set $k_v = 0.2$ so that $|G(j\omega)|$ is slightly larger than 1, as suggested by Fig. 8(a). Similarly, in the string stable case, $k_v = 1$ and $|G(j\omega)| \leq 1$.

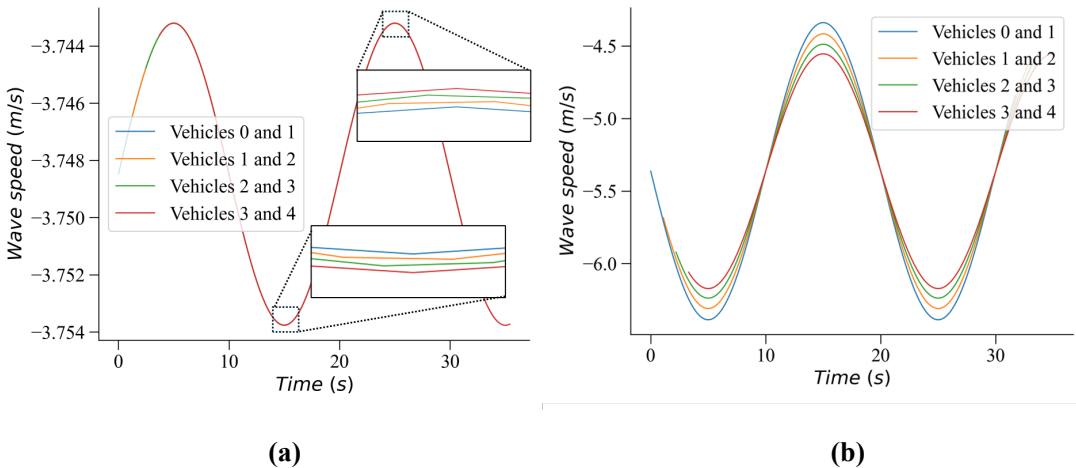

(a)                          (b)

**Fig. 11** Wave speeds over different $k_v$ (a) String instable $k_v = 0.2$; (b) String stable $k_v = 1.0$

It can be observed that the wave speed exhibits more pronounced fluctuations when the initial oscillation propagates to the end of the vehicle string, as indicated by its larger amplitude. In contrast, the string stable case shows that the wave speed has smaller amplitude after initial oscillations propagating to the end of the platoon, suggesting that string stability helps to reduce the wave speed oscillation.

**Impact on Nonlinearity**

We examine here the implications of oscillation amplitude under different boundary conditions on wave speed. To this end, the experiment sets boundaries on $\hat{v}$ as illustrated in Fig 3. The purpose here is to show how disturbances propagate along a vehicular string. Since the disturbance amplification ratio essentially depends on the ratio of the amplitude, $A$, of the external input (i.e., the amplitude of the leader's velocity oscillation) and $v_{bound}$, we examine the amplification ratio and phase shift under different $A/v_{bound}$. For simplicity and illustration purposes, we set $v_e = v_{free} - v_e = v_{bound} = 10$ And vary $A$ from 8 to 10. Note that, the nonlinear boundaries usually coexist with the string unstable case, so $\tau$ is adjusted to 0.5s to induce linear string instability. We set $\omega = 1Hz$ while other parameters follow those in Table 1. The detailed amplification ratio $|G(j\omega)|$, response time, as well as wave speed are given in Fig. 12. It is observed that the oscillation damping becomes more effective when the limits are reached in the case $A/v_{bound} = 1$. For cases $A/v_{bound} = 0.8$ and $A/v_{bound} = 0.9$, the limits have not been reached, resulting in the same $|G(j\omega)|$. As seen in Fig. 12(b), the response time remains insignificantly affected by the nonlinearity, which is anticipated since saturation itself is a nonlinear component that introduces zero phase shift to the traversing signal. As explained in Section 4.2, the wave speed oscillation is linearly proportional to $|1 - |G(j\omega)||$, which validate the result in Fig. 12(c) that, when $A/v_{bound}$ increases to $A/v_{bound} = 1$ from $A/v_{bound} = 0.9$, the wave speed oscillation amplitude becomes smaller. Also by Eq. (10), the wave speed oscillation is linearly proportional to $A$, given the cases $A/v_{bound} = 0.8$ and $A/v_{bound} = 0.9$ have the same amplification ratio, we can also validate that in the case $A/v_{bound} = 0.8$ v.s. $A/v_{bound} = 0.9$, the wave speed oscillation amplitude becomes larger as $A$ becomes larger.

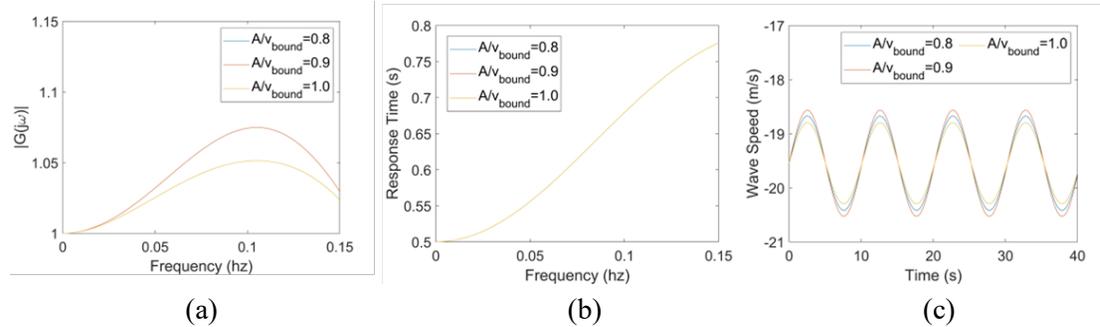

**Fig. 12** Behavior over different boundaries limit (a) Disturbance amplification ratio; (b) Response time; (c) wave speed

## CONCLUSION

In conclusion, this paper provides a significant contribution to the traffic flow theory by establishing a comprehensive theoretical understanding of traffic waves specific to Automated Vehicles (AVs). Our work focuses on understanding the relationship between AV longitudinal control and the fundamental properties of traffic waves, extending the existing knowledge that previously centered on (HDVs). It is evident from our study that traffic waves in AVs are markedly more complex, and their propagation is intricately tied to the control structure, control parameters, and the AV's response to traffic disturbances.

In congruence with the understanding of traffic wave propagation in human traffic, we attribute the propagation of traffic waves in AVs to their control paradigm. However, our study underscores the complexity that arises from the specific control logic, parameter settings, stability conditions, and mathematical properties of AVs' control systems. By deciphering how these factors influence the AV's response to different traffic disturbances, we provide substantial insights into the traffic waves generated by AVs and their fundamental characteristics.

Our theoretical derivation of traffic waves for homogeneous and heterogeneous platoons with both single and compounding frequency oscillations offers a robust and comprehensive framework for approximating and understanding AV traffic waves. Notably, the insights and properties we have developed provide a foundation for designing AV control systems that can provide desired properties on wave propagation. This has potential to improve traffic-level properties, such as string stability, opens exciting prospects for the development of more effective control strategies for AVs, ultimately contributing to a more efficient and safer transportation system.

The results of the numerical experiments presented in this paper serve as validation for our theoretical framework. However, we acknowledge that real-world applications may present complexities beyond our current understanding. Hence, we emphasize that continual research and exploration into this area is necessary, especially as AV technologies continue to evolve and influence our traffic systems. We hope that our research sparks further investigation and serves as a significant steppingstone in the broader quest to optimize our future transportation systems with the increasing presence of AVs.

Zhou, Li, Kontar, Pu, Srivastava, and Ahn*Part B: Methodological*, Vol. 36, No. 3, 2002, pp. 195–205. https://doi.org/10.1016/S0191-2615(00)00044-8.

17. Ahn, S., M. J. Cassidy, and J. Laval. Verification of a Simplified Car-Following Theory. *Transportation Research Part B: Methodological*, Vol. 38, No. 5, 2004, pp. 431–440. https://doi.org/10.1016/S0191-2615(03)00074-2.

18. Chen, D., J. Laval, Z. Zheng, and S. Ahn. A Behavioral Car-Following Model That Captures Traffic Oscillations. *Transportation Research Part B: Methodological*, Vol. 46, No. 6, 2012, pp. 744–761. https://doi.org/10.1016/j.trb.2012.01.009.

19. Chen, D., J. A. Laval, S. Ahn, and Z. Zheng. Microscopic Traffic Hysteresis in Traffic Oscillations: A Behavioral Perspective. *Transportation Research Part B: Methodological*, Vol. 46, No. 10, 2012, pp. 1440–1453. https://doi.org/10.1016/j.trb.2012.07.002.

20. Jiang, R., Q. Wu, and Z. Zhu. Full Velocity Difference Model for a Car-Following Theory. *Physical Review E*, Vol. 64, No. 1, 2001, p. 017101. https://doi.org/10.1103/PhysRevE.64.017101.

21. Laval, J. A., and L. Leclercq. A Mechanism to Describe the Formation and Propagation of Stop-and-Go Waves in Congested Freeway Traffic. *Philosophical Transactions of the Royal Society A: Mathematical, Physical and Engineering Sciences*, Vol. 368, No. 1928, 2010, pp. 4519–4541. https://doi.org/10.1098/rsta.2010.0138.

22. Treiber, M., and A. Kesting. An Open-Source Microscopic Traffic Simulator. *IEEE Intelligent Transportation Systems Magazine*, Vol. 2, No. 3, 2010, pp. 6–13.

23. Chiabaut, N., and L. Leclercq. Wave Velocity Estimation through Automatic Analysis of Cumulative Vehicle Count Curves. *Transportation Research Record*, Vol. 2249, No. 1, 2011, pp. 1–6. https://doi.org/10.3141/2249-01.

24. Taylor, J., X. Zhou, N. M. Rouphail, and R. J. Porter. Method for Investigating Intradriver Heterogeneity Using Vehicle Trajectory Data: A Dynamic Time Warping Approach. *Transportation Research Part B: Methodological*, Vol. 73, 2015, pp. 59–80. https://doi.org/10.1016/j.trb.2014.12.009.

25. Zheng, Z., S. Ahn, D. Chen, and J. Laval. Freeway Traffic Oscillations: Microscopic Analysis of Formations and Propagations Using Wavelet Transform. *Procedia - Social and Behavioral Sciences*, Vol. 17, 2011, pp. 702–716. https://doi.org/10.1016/j.sbspro.2011.04.540.

26. Tian, J., C. Zhu, D. Chen, R. Jiang, G. Wang, and Z. Gao. Car Following Behavioral Stochasticity Analysis and Modeling: Perspective from Wave Travel Time. *Transportation Research Part B: Methodological*, Vol. 143, 2021, pp. 160–176. https://doi.org/10.1016/j.trb.2020.11.008.

27. Zhou, H., A. Zhou, T. Li, D. Chen, S. Peeta, and J. Laval. Significance of Low-Level Controller for String Stability under Adaptive Cruise Control. http://arxiv.org/abs/2104.07726. Accessed Aug. 2, 2024.

28. Zhou, Y., X. Zhong, Q. Chen, S. Ahn, J. Jiang, and G. Jafarsalehi. Data-Driven Analysis for Disturbance Amplification in Car-Following Behavior of Automated Vehicles. *Transportation Research Part B: Methodological*, Vol. 174, 2023, p. 102768. https://doi.org/10.1016/j.trb.2023.05.005.

29. Jiang, J., Y. Zhou, X. Wang, and S. Ahn. On Dynamic Fundamental Diagrams: Implications for Automated Vehicles. *Transportation Research Part B: Methodological*, 2024, p. 102979. https://doi.org/10.1016/j.trb.2024.102979.
24